%% file: main.tex
\documentclass[letterpaper,twocolumn,10pt]{article}
\usepackage{usenix-2020-09}

\usepackage{tikz}
\usepackage{amsmath}
\usepackage{amsfonts}
\usepackage{makecell}
\usepackage{algorithm}
\usepackage{algpseudocode}
\usepackage{caption}
\usepackage{subcaption}
\usepackage{appendix}
\usepackage{xurl}
\usepackage{xspace} 
\usepackage{booktabs}
\PassOptionsToPackage{table}{xcolor}
\usepackage{titling}
\usepackage{lastpage}
\usepackage{arydshln}           % dashed lines for tables

\usepackage{listings}
\usepackage{xcolor}
\colorlet{punct}{red!60!black}
\definecolor{background}{HTML}{EEEEEE}
\definecolor{delim}{RGB}{20,105,176}
\colorlet{numb}{magenta!60!black}
\lstdefinelanguage{json}{
    basicstyle=\footnotesize\sffamily,
    numbers=left,
    numberstyle=\scriptsize,
    stepnumber=0,
    numbersep=8pt,
    showstringspaces=false,
    breaklines=true,
    frame=lines,
    backgroundcolor=\color{background},
    literate=
     *{0}{{{\color{numb}0}}}{1}
      {1}{{{\color{numb}1}}}{1}
      {2}{{{\color{numb}2}}}{1}
      {3}{{{\color{numb}3}}}{1}
      {4}{{{\color{numb}4}}}{1}
      {5}{{{\color{numb}5}}}{1}
      {6}{{{\color{numb}6}}}{1}
      {7}{{{\color{numb}7}}}{1}
      {8}{{{\color{numb}8}}}{1}
      {9}{{{\color{numb}9}}}{1}
      {:}{{{\color{punct}{:}}}}{1}
      {,}{{{\color{punct}{,}}}}{1}
      {\{}{{{\color{delim}{\{}}}}{1}
      {\}}{{{\color{delim}{\}}}}}{1}
      {[}{{{\color{delim}{[}}}}{1}
      {]}{{{\color{delim}{]}}}}{1},
}

\definecolor{codegreen}{rgb}{0,0.6,0}
\definecolor{codegray}{rgb}{0.5,0.5,0.5}
\definecolor{codepurple}{rgb}{0.58,0,0.82}
\definecolor{backcolour}{rgb}{0.95,0.95,0.92}
\lstdefinestyle{mystyle}{
    backgroundcolor=\color{backcolour},   
    commentstyle=\color{codegreen},
    keywordstyle=\color{magenta},
    numberstyle=\tiny\color{codegray},
    stringstyle=\color{codepurple},
    basicstyle=\ttfamily\scriptsize,
    breakatwhitespace=false,         
    breaklines=true,                 
    captionpos=b,                    
    keepspaces=true,                 
    numbers=left,                    
    numbersep=5pt,                  
    showspaces=false,                
    showstringspaces=false,
    showtabs=false,                  
    tabsize=2
}

\usepackage{amsmath}

\usepackage{filecontents}
\usepackage{tabularx}
\usepackage{pifont}

\usepackage{tikz}
 
\newcommand*\emptycirc[1][1ex]{\tikz\draw (0,0) circle (#1);} 
\newcommand*\halfcirc[1][1ex]{%
  \begin{tikzpicture}
  \draw[fill] (0,0)-- (90:#1) arc (90:270:#1) -- cycle ;
  \draw (0,0) circle (#1);
  \end{tikzpicture}}

\newcommand*\circled[1]{\tikz[baseline=(char.base)]{
            \node[shape=circle,draw,inner sep=0.5pt] (char) {#1};}}

\newcommand{\algoname}{incremental scaling algorithm}
\newcommand{\sysname}{Kivi\xspace}
\newcommand{\correct}{accuracy}
\newcommand{\domainname}{cluster management systems\xspace}
\newcommand{\dname}{cluster management system\xspace}

\providecommand{\mypara}[1]{\smallskip\noindent\emph{#1} }
\providecommand{\myparab}[1]{\smallskip\noindent\textbf{#1} }

\newcolumntype{P}[1]{>{\centering\arraybackslash}p{#1}}

\newcommand{\totalfc}{16}

\newcommand{\hone}{C1}
\newcommand{\htwo}{C2}
\newcommand{\sthree}{C3}
\newcommand{\sfour}{C4}
\newcommand{\sone}{C5}
\newcommand{\ssix}{C6}
\newcommand{\done}{C7}
\newcommand{\snine}{C8}

\newcommand{\pplacement}{Unexpected Topology}
\newcommand{\pnumber}{Unexpected Object Numbers}
\newcommand{\plifecycle}{Unexpected Object Lifecycles}

\newcommand{\rcon}{Non-trivial interactions between components in a single controller}
\newcommand{\rcontroller}{Non-trivial interactions between controllers}
\newcommand{\revents}{Non-trivial interactions between controllers and events}
\usepackage{enumitem}
\setlist[enumerate]{itemsep=0mm}

\begin{document}
%-------------------------------------------------------------------------------

%don't want date printed
\date{}
\setlength{\droptitle}{-4em}   % Eliminate vertical space above title
\posttitle{\par\end{center}}   % tighten space between title and author block

\title{\Large \textbf{Kivi: Verification for Cluster Management} \\ \begin{large} \emph{(Draft under submission)} \end{large}}

%for single author (just remove % characters)
\author{
{\rm Bingzhe Liu}\\
UIUC
\and
{\rm Gangmuk Lim}\\
UIUC
% copy the following lines to add more authors
\and
{\rm Ryan Beckett}\\
Microsoft
\and
{\rm 
P. Brighten Godfrey}\\
UIUC and VMware
} % end author

\maketitle

%-------------------------------------------------------------------------------

\begin{abstract}
%-------------------------------------------------------------------------------

Modern cloud infrastructure is powered by \domainname{} such as Kubernetes and Docker Swarm. While these systems seek to minimize users' operational burden, the complex, dynamic, and non-deterministic nature
of these systems makes them hard to reason about, potentially leading to failures ranging from performance degradation to outages.

We present \sysname, the first system for verifying controllers and their configurations in \domainname. 
\sysname{} focuses on the popular system Kubernetes, and models its controllers and events into processes whereby their interleavings are exhaustively checked via model checking. Central to handling autoscaling and large-scale deployments is our design that seeks to find violations in a smaller and reduced topology. We also develop several model optimizations in \sysname{} to scale to large clusters. We show that \sysname{} is effective and accurate in finding issues in realistic and complex scenarios and showcase two new issues in Kubernetes controller source code. 
\end{abstract}

\input{intro}
\input{background}
\input{failure_cases}

\input{system_arch}
\input{modeling}

\input{eval}

\input{discussion}

\input{related_work}
\input{conclusion}
\bibliographystyle{plain}
\bibliography{ref}
\input{appendix}

%%%%%%%%%%%%%%%%%%%%%%%%%%%%%%%%%%%%%%%%%%%%%%%%%%%%%%%%%%%%%%%%%%%%%%%%%%%%%%%%
\end{document}

%% file: intro.tex
%-------------------------------------------------------------------------------
\vspace{-1em}
\section{Introduction}
\vspace{-0.5em}

Modern cloud infrastructure relies on technologies like containers, microservices, and serverless computing to develop applications that are resilient and manageable. To orchestrate these components, practitioners are widely adopting cluster management systems like Kubernetes and Docker Swarm. These \domainname{} consist of a diverse collection of controllers (e.g., scheduler, autoscaler, deployment controller). Aiming to minimize the users' operational burden, these controllers run in a continuous closed loop and automatically drive the cluster towards desired goals through manipulating shared cluster resources. 

However, it is challenging for users to safely and correctly use these \domainname{}, and many outages have occurred as a result~\cite{outage_1, outage_2, s4, syntax1, outage_4}. Even a single controller can consist of sophisticated workflow logic that is influenced by diverse configuration parameters, introducing opportunities for human mistakes. For example, the scheduler in Kubernetes has 12 different pipeline stages and 21 complex scheduling strategy plugins (each of which contains a few or even a dozen of parameters to tune) to choose from for these stages~\cite{scheduler_config}. Moreover, even if users correctly configure a single controller, multiple controllers can have non-trivial interactions because they operate on overlapping sets of cluster components~\cite{liu2020towards}. Furthermore, frequent operational changes, multiple teams of engineers making changes with different goals, and unpredictable environmental events (e.g., workload changes and failures) all amplify the chance of a mistake.

Through case studies of community-collected failure cases~\cite{failure_stories}, we have identified a collection of unintended or pathological behaviors (\S\ref{sec:motivating_examples}) that can happen due to the aforementioned challenges. Unintended behaviors include the number of pods dropping below a necessary minimum, or the placement of pods becoming unexpectedly unbalanced. Pathological behaviors include a pod oscillating in an unending cycle of scheduling and eviction, or failing to be scheduled despite the existence of plentiful resources. These behaviors can result in performance degradation and even serious service outages, yet are hard for users to reason about manually. 

With the increasing complexity of \domainname{}, we believe users need automated ways to check the correctness of their clusters. Testing and emulation~\cite{quickcheck, DBLP:conf/osdi/YuanLZRZZJS14, flymc, crystalnet, softwareupgradefailure, ctest, heteroconfig, acto, morpheus} are common ways to find issues in any system, but they are insufficient for \domainname where controllers run asynchronously 
%in a distributed fashion 
and events may take place in any non-deterministic order. On the other hand, verification techniques are known to provide high coverage of distributed systems by considering all interleavings of components. There have been many success stories using verification for distributed systems and networking (e.g., \cite{DBLP:conf/osdi/HanceLHHJP20, DBLP:conf/osdi/Sigurbjarnarson18, DBLP:conf/sosp/ChajedTKZ19, ironfleet, rehearsal, DBLP:conf/sosp/ZouDDFGC19, DBLP:conf/osdi/YaseenABCL20, DBLP:conf/sosp/BornholtJACKMSS21, DBLP:conf/sosp/TaoYLLNG21, liveness, minesweeper, kinetic, netsmc, zhang_et_al:LIPIcs.ITP.2021.32}), yet none have been tailored to \domainname. They either focus on low-level implementation details or specific protocols (e.g., Paxos, BGP), instead of the properties that are related to the entanglement of several or even tens of control components. For instance, in Kubernetes, the number of replicas is affected by the deployment controller, horizontal pod autoscaling (HPA), scheduler (and its various plugins), de-scheduler, events like node failures, and more. 

In this paper, we present \sysname{}, the first system for verifying \dname controllers and their configurations. \sysname{} focuses on Kubernetes, the most popular open-source cluster management platform. \sysname{} takes the users' intent, the state and configuration of the cluster, and the event assumptions as inputs, and verifies if the cluster can violate the intent. If a violation is possible, it generates a minimal counterexample.

Our main goal in \sysname{} is to verify the \emph{interactions} between controllers or between controllers and events. Therefore, instead of verifying the large-scale source code of Kubernetes, we focus on modeling the high-level logic of the controllers that carry out the essential logic affecting the shared global states (i.e., the status of pods and nodes). \sysname{} leverages the explicit model checker SPIN~\cite{spin}. We model Kubernetes objects into shared global states and model controllers and events into processes such that their non-deterministic interleaving can be searched exhaustively by SPIN. 

Scalability is a daunting challenge in \domainname. A cluster can reach hundreds of nodes and many thousands of pods leading to the state explosion problem~\cite{state_explosion} in verification. Furthermore, autoscalers are common in \domainname, so users would be interested in not only one but a wide range of cluster topologies. There may be multiple dimensions for autoscaling (i.e., various types of nodes and pods), and the possible topologies that need to be verified can grow exponentially against the size of the cluster. 

To tackle the scalability challenges, we posit a hypothesis: \emph{if a cluster setup can violate an intent, then it can do so at relatively small scale}. Prior work~\cite{DBLP:conf/osdi/YuanLZRZZJS14} has made a similar observation in other types of distributed systems. According to this intuition, \sysname{} is designed with an \algoname{}, where it starts to verify a cluster at the smallest nontrivial scale, and then intelligently increases the scale of the system across multiple scaling dimensions until either finding a violation or reaching a scale threshold that is large enough to conclude, with high experimental confidence, that no violation is possible at any scale. Although this approach does not provide the absolute guarantee of a fully exhaustive exploration, it dramatically improves performance compared to testing at all possible cluster scale, and also produces violations that are more minimal and easier for users to understand. 
In addition, to improve the performance of individual runs of the model, we have implemented a few optimization mechanisms to reduce the verification search space, which enables our model to scale to sufficiently large problem sizes.

% shortend
\sysname{} implements six commonly used and representative Kubernetes controllers with logic derived from the Kubernetes source code. We evaluate \sysname{} on a test suite of eight representative cases derived from realistic failures. We summarize the key findings as follows.
\vspace{-0.05in}
\begin{itemize}[leftmargin=*]
   \itemsep0em 
   \item \emph{Validating the intuition.} Using \sysname{}, we show that intent violations consistently appear at small scale: the largest minimum size needed to produce a violation is only 3 nodes and 6 pods. Indeed, in 6 of 7 cases, \sysname{} found violations at even smaller scale than the original problem report.
   %, and we confirmed these by running the configurations in a real Kubernetes cluster.
   
   \item \emph{Performance.} Our evaluation using realistic failure cases shows that \sysname verifies most cases within 100 seconds and all cases within 25 minutes.  Without our \algoname{}, verification times out ($>10000$ sec) even at moderate scale ($\leq 50$ nodes).
   
   \item \emph{Accuracy.} \sysname{} has successfully found the correct violations for all configuration violation cases and reported no failures for non-violation cases. We have also performed a comparison with real Kubernetes cluster runs and found that \sysname closely models the real system.
   
   \item \emph{New issues found.} \sysname{} has found two new issues in a Kubernetes controller.
\end{itemize}
\vspace{-0.05in}

%% file: background.tex
\vspace{-1.5em}
\section{Modern Cluster Management Systems} \label{sec:background}
\vspace{-1em}

Cluster management systems~\cite{kubernetes, twine, dockerswarm, vsphere} orchestrate and manage the lifecycles of compute resources by allocating and scaling resources efficiently to improve application performance and reliability.
These cluster management systems consist of multiple controllers that operate asynchronously to drive the cluster towards desired goals through manipulating shared objects (a.k.a. workload resources). While the concepts are more general, we describe here the basic moving parts in Kubernetes. \textbf{\emph{Pods}}\footnote{Pods are also called replicas in the Kubernetes deployment configuration. We use replicas and pods interchangeably in this paper.} are the smallest deployable unit.  Each pod consists of one or several containers with shared resources. \textbf{\emph{Nodes}} are virtual or physical machines that run pods. \textbf{\emph{Deployments}} specify information about the lifecycle of a group of pods -- including both user configuration guiding how they should be scaled, updated, and terminated, and also the live state such as number of replicas.

Kubernetes provides many controllers that manage the above resources.
Each controller contains \emph{reconciliation} logic that attempts to drive the cluster state back to its goal state by periodically checking the cluster state via the logically centralized communication channel \textit{API server}. Though they share states via the API server, there is no agreed-upon notion of a globally desired goal state. We now briefly explain the goal of a few key Kubernetes controllers that are covered later in the paper. \textit{\textbf{Scheduler}} places pods to nodes. It filters out infeasible nodes and selects the best node based on scheduling preference. The scheduler can be configured with different plugins, e.g., \texttt{PodTopologySpread} is used to evenly distribute a group of pods to prevent skewed pod placement between nodes, and \texttt{NodeAffinity} schedules pods onto preferred nodes.
% removeduplicated / removepodspreading
\textit{\textbf{Descheduler}} is an additional controller to actively evict pods when the pod placement could drift away from the desired state.\footnote{Note that the scheduler does not move pods dynamically once placed.}
% It has multiple plugins user can configure for one's circumstances. 
For example, the \texttt{RemoveDuplicate} plugin evicts pods when there are two identical pods replicated on the same node.
% but mention HPA as the major example 
\textit{\textbf{Horizontal Pod Autoscaler (HPA)}} is a controller that autoscales the number of pods based on the given target resource utilization (e.g., CPU usage) within specified minimum and maximum limits. 
% maxSurge, ReCreate, RollingUpdates
\textit{\textbf{Deployment and Replicaset Controllers}} implement the logic to manage the lifecycle of pods, guided by the deployment specification.
\textit{\textbf{Kubelet}} is a controller residing on each node. It is responsible for managing the lifecycle of pods in each node. 

%% file: failure_cases.tex
\vspace{-1.2em}
\section{Motivating Examples and Takeaways for Verification}
\vspace{-.5em}

We studied community-collected failure cases~\cite{failure_stories} supplemented with failure cases mentioned in various talks at the main conference for the Kubernetes developer and user community KubeCon 2017-2023 and potential problems mentioned in the Kubernetes official documentation. While there are various reasons for failures, including DNS issues~\cite{dns1, dns2}, Linux kernel issues~\cite{linux1, linux2, linux3}, configuration syntax problems~\cite{syntax1, syntax2}, and credential issues~\cite{credential1, syntax2}, we are interested in the failures that are caused by the non-trivial interactions between controllers, or between controllers and events. We have collected \totalfc{} failure cases related to the non-trivial interactions and summarized them in Table~\ref{tab:failure_case_1} and an extended Table~\ref{tab:failure_case_2} in the Appendix~\ref{sec:more_failure}. 

In \S\ref{sec:motivating_examples}, we summarize the causes of failure into three categories and describe a few motivating examples. Based on these failures we identify several important properties to verify for \domainname{} in \S\ref{sec:properties}. We finally discuss the takeaways for verification tool design in \S\ref{sec:takeaways}.

\begin{table*}[h]
\renewcommand*{\arraystretch}{1}
\begin{center}
\scalebox{0.78}{
\begin{tabular}{ P{0.45in}P{3.25in}P{1.7in}P{.6in}P{0.88in}P{0.95in} }
\toprule
\textbf{Case ID} & \textbf{Description} & \textbf{Properties} & \textbf{Reasons} & \textbf{Min Violation Scale} $\langle|N|, |P|\rangle$ & \textbf{Reported Scale} \newline $\langle|N|, |P|\rangle$ \\
\midrule

\hone~\cite{h1} & Pods consumed high CPU during bootstrapping leading HPA to scale up rapidly to max replicas. & Unexpected object numbers & CE
%HPA \newline Pod resource usage change 
& $\langle1, 3\rangle$ & $\langle \mathrm{Unknown}, 150+\rangle$ \\
\addlinespace[.3ex]
\hdashline[1pt/1pt]
\addlinespace[.3ex]
\htwo\cite{h2, h2_1} & Not enough replicas because users apply an updated YAML file without defining number of replicas (1 by default). & Unexpected object numbers & MC, CE
%HPA \newline Deployment Controller \newline User command "kubectl apply" 
& $\langle 1, 2\rangle$ & $\langle \mathrm{Unknown}, 3\rangle$ \\
\addlinespace[.3ex]
\hdashline[1pt/1pt]
\addlinespace[.3ex]
\sthree~\cite{s3} & Configurations of two \texttt{PodTopologySpread} constraints caused the 6th pod to fail to be scheduled. & Unexpected object lifecycles & SCC
%HPA \newline Scheduler (PodTopoSpread) 
& $\langle3, 6\rangle$ & $\langle3, 6\rangle$\\
\addlinespace[.3ex]
\hdashline[1pt/1pt]
\addlinespace[.3ex]
\sfour~\cite{s4} & Scheduler kept assigning pods with high CPU usage to the same node causing a kernel panic and pod failure loop. & Oscillation / Unexpected object lifecycles & CE
%Scheduler \newline Kernel panic attack 
& $\langle2, 5\rangle$ &  \textrm{Unknown} \\
\addlinespace[.3ex]
\hdashline[1pt/1pt]
\addlinespace[.3ex]
\sone~\cite{s9} & Conflict configurations of scheduler and \texttt{RemoveDuplicate} policy in descheduler. & Oscillation & MC
%Descheduler (RemoveDuplicates) \newline Scheduler (NodeResourceFit) 
& $\langle2, 4\rangle$ &  $\langle5, 10\rangle$ \\
\addlinespace[.3ex]
\hdashline[1pt/1pt]
\addlinespace[.3ex]
\ssix~\cite{s6} & Pod distribution unbalanced after maintenance. Node failures then caused the pod count to drop too low. & Unexpected object placement / numbers & CE
%\newline numbers & Scheduler (PodTopoSpread) \newline Maintenance 
& $\langle 2, 2\rangle$ & $\langle3, 3\rangle$ \\
\addlinespace[.3ex]
\hdashline[1pt/1pt]
\addlinespace[.3ex]
\done~\cite{d1} & Conflicting configurations of node \texttt{taint} and pod \texttt{NodeName} caused scheduling and eviction loop & Oscillation / Unexpected object lifecycles & MC
%Taint manager/Node Controller \newline Scheduler (TaintToleration) 
& $\langle 1, 1\rangle$ &  $\langle \mathrm{Unknown}, 5\rangle$ \\
\addlinespace[.3ex]
\hdashline[1pt/1pt]
\addlinespace[.3ex]
\snine~\cite{s9} & Conflicting descheduler and scheduler configurations caused scheduling and eviction loop. & Oscillation / Unexpected object lifecycles & MC
%Descheduler (RemoveViolatingTopoSpread) \newline Scheduler (NodeAffinity, PodTopoSpread) 
& $\langle3, 6\rangle$  &  $\langle 5, 10+\rangle$ \\
\bottomrule
\end{tabular}
}
\end{center}
\vspace{-1.8em}
\caption{\label{tab:failure_case_1} \small Failure cases that are caused by the non-trivial interactions between control components and between controllers and events. The last two column shows the minimum scale that we find violations for each case, and whether it is smaller than the reported scale. MC stands for nontrivial interactions between \emph{multiple controllers}, SCC for \emph{single controller components}, an CE for \emph{controllers and events}.
\vspace{-1.5em}
}
\end{table*}

\vspace{-1em}
\subsection{Why do problems occur?} \label{sec:motivating_examples}
\vspace{-0.5em}
\myparab{\rcon.} Even configuring a single controller correctly is already challenging. For example, the scheduler contains 12 pipeline stages that can be extended with a total of 21 default plugins~\cite{scheduler_config} for a wide range of scheduling strategies. Each of these plugins contains various configurations, some of which can interact with other plugins. For example, users can configure the \texttt{PodTopologySpread(SPTS)} with an unlimited number of constraints where each constraint has 8 parameters, 2 of which are related to another two plugins~\footnote{The configurations define if \texttt{NodeAffinity} and \texttt{NodeTaints} would be considered when calculating the skewness in topology.}. Furthermore, many important details have not been documented well and users can easily make mistakes. For example, if multiple \texttt{SPTS} constraints are defined for a pod, the node's labels need to match with all constraints to be considered in skewness calculation and as a candidate for the pod.

\begin{figure}
     \centering
     \begin{subfigure}[b]{0.165\textwidth}
         \centering
         \includegraphics[width=\textwidth]{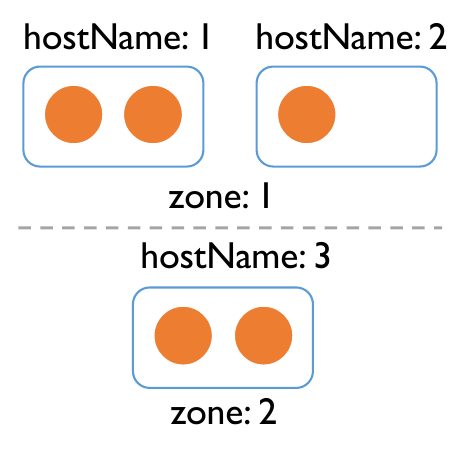}
          \vspace{-1.7em}
         \caption{\small Case \sthree}
         \label{fig:case_s3_topo}
     \end{subfigure}
     \hfill  \hspace{0.01in} 
     \begin{subfigure}[b]{0.3\textwidth}
         \centering
         \includegraphics[width=\textwidth]{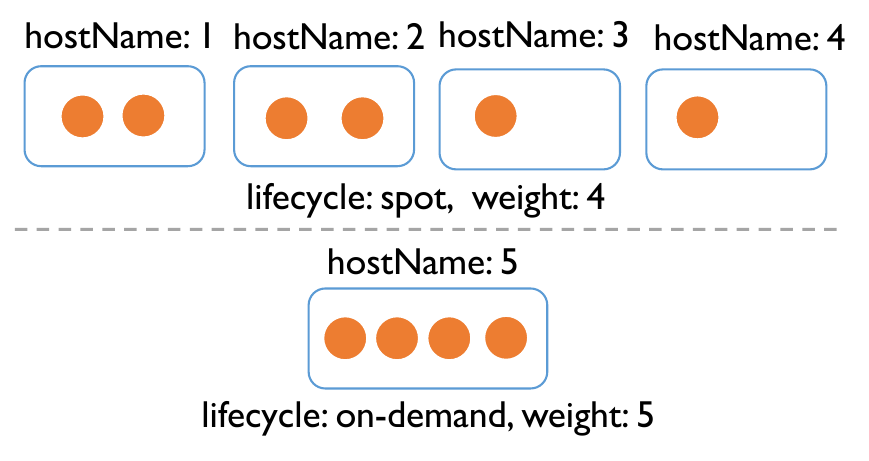}
          \vspace{-1.7em}
         \caption{\small Case \snine}
         \label{fig:case_s9_topo}
     \end{subfigure}
     \vspace{-2em}
     \caption{\small Cluster configurations for two failure examples.}
        \label{fig:topo}
    \vspace{-1.5em}
\end{figure}

We describe a simple example that shows the non-trivial interaction between configurations in a single plugin. Before we dive into details, we first simplify and define each \texttt{SPTS} constraint as \texttt{skew(}$k$\texttt{)$\leq$}$a$, which means the pods need to be approximately evenly spread onto the groups of nodes identified by having the same value of label $k$ such that the difference in the number of pods in any two groups is $\leq a$. For example, the constraint \texttt{skew(zone)$\leq$ 1} is satisfied by Fig.~\ref{fig:case_s3_topo}: there are two groups, one defined by the label \texttt{zone:1} having 3 pods, and the other defined by \texttt{zone:2} and having 2 pods, and hence the difference between the two groups is $1$.

\mypara{Case \sthree{} (Conflict SPTS constraints)}\cite{s3}. A 3-node cluster was labeled as shown in Fig.~\ref{fig:case_s3_topo}. A 6-pod deployment was configured with two \texttt{SPTS} constraints: \texttt{skew(hostName)$\leq$ 1} and \texttt{skew(zone)$\leq$ 1}. The first 5 pods were scheduled successfully as shown in the figure. However, the 6th pod failed to be scheduled because it could not satisfy both constraints. This issue only manifests under specific topologies. If each zone is configured with a similar number of nodes, the user would not encounter this issue.

\myparab{\rcontroller.} Kubernetes contains various controllers, each of which has its own goals while manipulating shared objects. Furthermore, these controllers may be configured by multiple teams with different goals. The interactions between these controllers may affect each other's optimization goals or even become conflicts, causing performance degradation or even pathological behavior. We illustrate this issue with two cases.

\mypara{Case \snine{} (Conflict between scheduler and descheduler)}\cite{s9}. A 5-node cluster was labeled with  \texttt{hostName} and \texttt{lifecycle} as shown in Fig.~\ref{fig:case_s9_topo}. A 10-pod deployment was configured to use two scheduling plugins: (1) \texttt{SPTS} with two soft constriants \texttt{skew(hostName)$\leq$1} and \texttt{skew(lifecycle)$\leq$1}; (2) \texttt{NodeAffinity(SNA)} where the nodes with \texttt{on-demand} for the label \texttt{lifecycle} are preferred during scheduling than the nodes with the value \texttt{spot}, with weight of 5 and 4 respectively.
The two \texttt{SPTS} constraints in fact conflict with each other (similar to Case \sthree) and with the \texttt{SNA}: \texttt{SPTS} is targeted to spread evenly while \texttt{SNA} prefers the \texttt{on-demand} nodes. Therefore, the \texttt{SPTS} constraints cannot be satisfied, yet scheduling was still successful as shown in the figure because \texttt{SPTS} were soft constraints. %According to these configs, the pods were created by the deployment controller and were scheduled by the scheduler as shown in the figure. 
However, a descheduler was configured to hold the \texttt{SPTS} constraints even though they were soft. The non-trivial interactions between three controllers led to an unending eviction and scheduling cycle: the descheduler evicted pods, the deployment controller added the pods back to maintain the desired replicas, and scheduler scheduled pods back.

\myparab{\revents.} Various events can occur that affect the status of the objects in a live cluster. For example, environmental events like pod CPU changes (e.g., due to increased user requests) and node failures, and operational events like maintenance and application deployment. Users may not consider these events when they configure controllers, and moreover, these events can happen non-deterministically, resulting in failures that are hard to predict prior to deployment.

\mypara{Case \ssix{} (Pod unbalance after maintenance)}\cite{s6}. Pods initially were evenly spread across the nodes. A maintenance event happened that took down a node, resulting in rescheduling of the affected pods. After the maintenance was completed and the node came back, however, the pods was not rescheduled back to the node, resulting in an unbalanced topology and leaving the cluster with potential vulnerability to failures. A descheduler should have been used to re-balance the pods.
%and increase the resilience. 

\mypara{Case \htwo{} (Exceeded number of pods caused by CPU usage)}\cite{h2}. The pods in a deployment consumed high CPU ($100\%$ of requested) at the bootstrapping phase and then dropped back to normal usage. However, the high CPU usage at the beginning led the HPA to scale up the pods rapidly until the maximum number of replicas allowed. After pods gradually went through the bootstrapping phase, the HPA then slowly scaled down these extra pods. These extra pods wasted many resources without serving actual traffic and caused great confusion for the users.

\vspace{-1em}
\subsection{Properties to Verify} \label{sec:properties} 
From the example failure cases, we have identified a few important properties to check in \domainname\footnote{These properties are not fully orthogonal. For example, Case \snine{} counts for both oscillation and unexpected object lifecycles. In these cases, the user can choose to verify either property.}. Failure cases are labeled with these properties in Table~\ref{tab:failure_case_1} and Table~\ref{tab:failure_case_2}. Among these properties, some are \emph{pathological} behaviors like the system state oscillating back and forth, and some are \emph{unintended} behavior that does not match with users' expectations. Violating these properties can result in sub-optimal resource utilization (and hence potential higher financial cost), less resilience to event disturbance, performance degradation, or even service outages. 

\myparab{Oscillation.} The state of the cluster can become unstable and change back and forth in an unending cycle. Case \snine{} shows an example when scheduling and evicting pods.

\myparab{\pplacement.} Objects should be placed in certain patterns according to users' optimization metrics. For example, user requests or pods may need to be evenly spread to improve failure tolerance, or certain nodes may contain special resources (e.g., GPUs) that should be saved for specific pods. Case \ssix{} demonstrates the example of unbalanced pods. 

\myparab{\pnumber.} The number of objects should fall into a certain range. Too many pods or nodes can consume excessive resources and too few can affect the reliability of the applications. Case \htwo{} illustrates the example of allocating pods excessively.

\myparab{\plifecycle.} The lifecycle of an object includes its creation, execution, and termination after task completion or failure. An object may go through an unexpected lifecycle, e.g., due to failure during creation or unexpected eviction. Pods fail to run stably in Case \snine{}.

\vspace{-1em}
\subsection{Takeaways} \label{sec:takeaways}
The goal of our case study is to learn the characteristics of the failures that are caused by the non-trivial interactions between components in the \domainname{}. Although the study is not broad enough to serve as a complete quantitative study for the domain, we believe the cases we collected have good diversity and are representative in terms of the involved Kubernetes components, the root causes, and the underlying properties being violated. These findings can guide us on what to model in \sysname{}: we should include events, controllers and their features (e.g., we should model different plugins in the scheduler), objects and their attributes (e.g., we should model the CPU usage of pods), properties mention in \S\ref{sec:properties}, and model quantities (both real and integer numbers).

We believe the lessons we learned can also help with other future verification work for \domainname{}.

%% file: system_arch.tex
\begin{figure*}[t]
\centering
\includegraphics[width=6.5in]{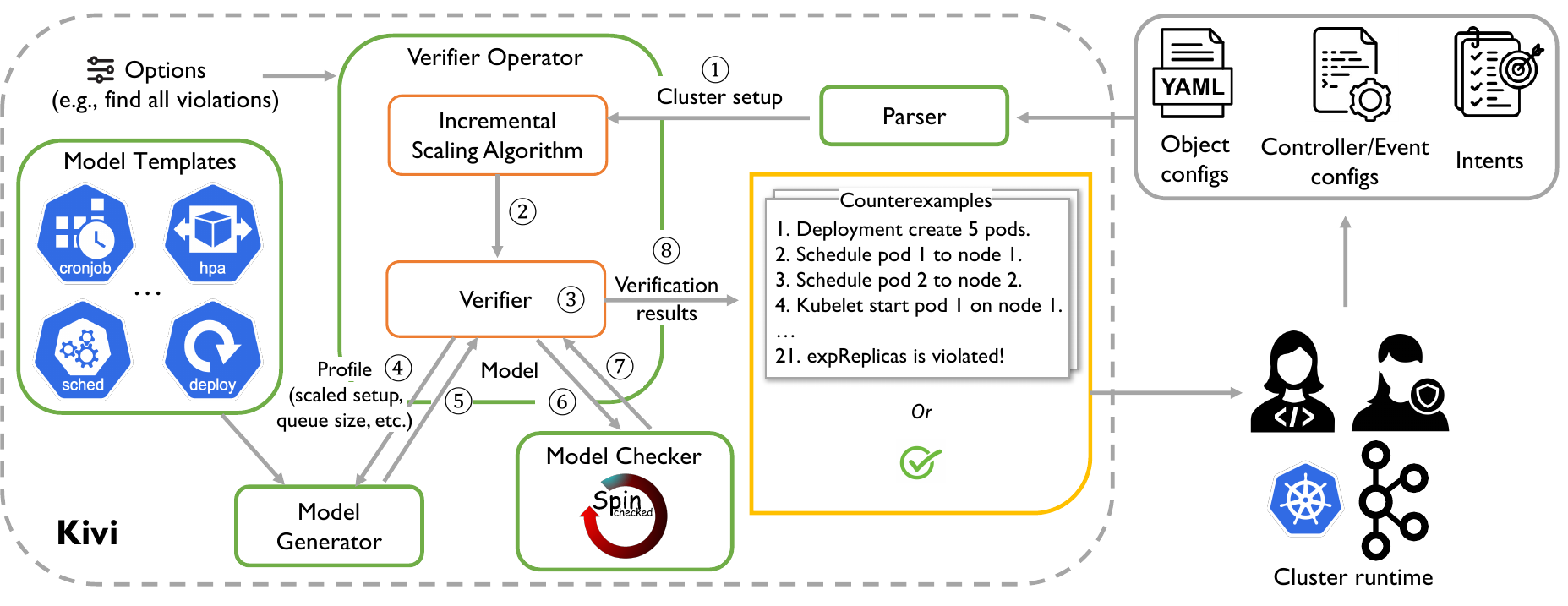}
\vspace{-0.5em}
\caption{\small\sysname{} workflow. It consists of five main components: \emph{Parser}, \emph{Verifier Operator}, \emph{Model Generator}, \emph{Model Templates} and \emph{Model Checker}. It receives the configurations and intents from the users and outputs verification results (i.e., counterexamples).}
\label{fig:veri_sys_arch}
\vspace{-1.5em}
\end{figure*}

\vspace{-1.2em}
\section{\sysname{} System Design}
\vspace{-1em}

To use \sysname, users provide their cluster configurations and select the intents that they want to verify from our property library. \sysname{} takes users' input, and either assures that the desired properties will be preserved or generates minimal counterexamples. \S\ref{sec:sys_overview} describes the workflow of \sysname{}.
A critical component of \sysname{} is the \emph{model}.  \sysname{} implements a carefully-designed model of Kubernetes controllers, objects, and events suitable for exhaustive analysis by a model checker.  In order to give this component a deep enough discussion, we describe it separately in \S\ref{sec:model}.

Verifying the \domainname{} raises significant performance challenges due to many execution paths, parameter configurations, and potentially large scale; we describe how Kivi tackles the problems of performance and scale with a scaling algorithm in \S\ref{sec:small_scale} and model design and optimization in \S\ref{sec:model}. We describe the workflow of the main component that operates the verification procedures in \S\ref{sec:verifier_operator}.

\vspace{-1em}
\subsection{System Workflow} \label{sec:sys_overview}
\vspace{-0.2em}
% may miss things about how do we expect users to use our system? 
 Figure~\ref{fig:veri_sys_arch} shows the workflow of \sysname, including the five main components of its design:

The \emph{Parser} takes three main elements as inputs: (1) the \emph{Object Configurations} that contains the configurations of nodes and Kubernetes workload resources, e.g., deployment YAML file, (2) the \emph{Controller and Event Configurations} that configure the behavior of the controllers and event assumption, e.g., the YAML or command lines of HPA and maintenance events on all the nodes, and (3) \emph{Intents} that describe operators' expected behavior of the cluster.
\emph{Parser} takes these inputs and parses them into a uniform format, the \emph{Cluster Setup} (more details in \S\ref{sec:alg}), and sends it to the \emph{Verifier Operator}. 

The \emph{Verifier Operator} carries out the actual verification procedure given the inputs from the \emph{Parser}. 
The verifier operator implements a scaling algorithm (introduced in \S\ref{sec:alg}) that conducts the verification in multiple \emph{cycles} with incrementally larger and larger scale. For each cycle, the verifier operator first decides the \emph{Profile}, which contains the setup of a particular scale, a subset of intents, and other verification configurations for that cycle. It then triggers the \emph{Model Generator} to generate a verification model of related controllers, events and cluster objects (more detail in \S\ref{sec:model}). It then receives the verification result from the \emph{Model Checker}, and finally decides if the results are ready to send to the users or if another circle should be triggered. We describe the workflow of the verifier operator in more detail in \S\ref{sec:verifier_operator}.

\begin{figure}[t]
\centering
\includegraphics[width=3.25in]{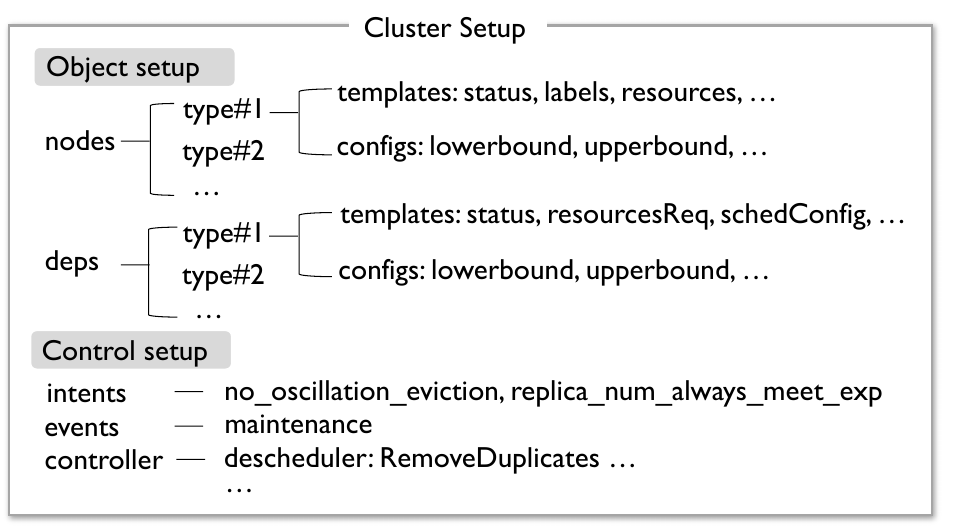}
\vspace{-1em}
\caption{\small Example of a cluster setup. It consists of the \emph{object setup} that defines the configurations for types of objects and the \emph{control setup} that includes the configurations for controllers, events and users' intents.}
\label{fig:cluster_setup}
\vspace{-1.8em}
\end{figure}

The \emph{Model Generator} pre-processes the \emph{Profile} to complete any missing elements with default values, discretizes all the values, and selects a subset of controller and event templates from the \emph{Model Templates} that are related to the designated intents and configurations according to the \emph{Profile}.

The \emph{Model Templates} include the model logic with ``holes'' that will be filled in with configuration parameters from the \emph{Profile} for the controllers and events \sysname{} supports. We have implemented these templates manually, where most controllers are modeled based on the Kubernetes source code. These independent templates enable the modularization of the verification process. We discuss more details in \S\ref{sec:model}.

The \emph{Model Checker} performs exhaustive verification for the model generated by the \emph{Model Generator}. It either assures desired properties will be preserved, or finds property violations and generates counterexamples demonstrating violations. We leverage the explicit-state model checker SPIN~\cite{spin} in \sysname{}.

\vspace{-1em}
\subsection{Verifying Clusters at Small Scale (But not too Small)}
% \subsection{Verifying Clusters at Small Scale (But not too Small)}
\vspace{-0.3em}
\label{sec:small_scale}

\subsubsection{Terminology}\label{sec:terminology}
\vspace{-0.5em}

We define a \emph{cluster setup} $C = \langle CS, OT \rangle$ as the configuration for the cluster. It consists of two main components, the \emph{control setup} $CS$ that includes the configurations for controllers, event assumptions, and users' intents, and the \emph{object setup} $OT$ that defines the configuration for types of objects, including their templates and replica ranges. \emph{Object setup} $OT = \langle NT, DT\rangle$,\footnote{It can also include other workload resources like StatefulSet, CronJob. We only discuss the Deployment for simplicity. } where $NT$ denotes the set of node types and $DT$ denotes the set of deployment types. Each node type $NT_i=\langle template, config \rangle$ consists of a template of object attributes (e.g., status, labels) and configs that define the lower bound $o_i^l$ and upper bound $o_i^u$ for the number of object type $o_i$. $DT$ is similar to $NT$ yet their templates contain different sets of attributes (e.g., resource requests, the scheduling configs). Figure~\ref{fig:cluster_setup} illustrates a graphic example of \emph{cluster setup}. \sysname{} derives the \emph{cluster setup} from users' inputs.

Each \emph{cluster setup} $C$ can generate a set of \emph{cluster topologies} $T^C$ at run time. We define a \emph{cluster topology} $t \in T^C$ as $t = \langle N, P\rangle$. $N$ is the set of nodes $N=N_1 \cup N_2 \cup \cdots$, where $N_i$ denotes the set of nodes of type $i$. $P$ is the set of pods $P= P_1\cup P_2 \cup \cdots$, where $P_i$ denotes the set of pods generated from the deployment of type $i$.

The scale of a cluster topology $t$ can be defined as a vector of the sizes for all types of objects $s_t = \langle |N_1|, |N_2|, \cdots, |N_k|, |P_1|, |P_2|, \cdots, |P_m|\rangle$, where $k$ and $m$ are the number of node types and deployment types respectively.

We define a \emph{scaled setup} $C_t = \langle t, C\rangle$ as an instance of the cluster setup $C$ running with topology $t$. 

\vspace{-1em}
\subsubsection{Incremental Scaling Algorithm Intuition}\label{sec:small_scale_insight}
\vspace{-0.5em}
On one hand, verifying \domainname{} has a daunting scale challenge.  Clusters can reach hundreds or thousands of nodes, and many thousands of pods, bringing well-known state explosion problems~\cite{state_explosion}.  Moreover, clusters typically run autoscalers and could vary in size, so that users would be interested in not only one but a wide range of the cluster topologies that are generated from a cluster setup, and there may be multiple dimensions of scale (see $s_t$ from \S\ref{sec:terminology}) so there is a very large number of possible cluster topologies even for a medium-sized cluster (tens of nodes).

On the other hand, there is a way in which the problem is actually simpler: much of the complexity mainly lies in the cluster setup which does not grow with the size of the topology. All the cluster topologies are generated from the cluster setup, where each object is generated from a template in the setup rather than being crafted by hand. Because of this, the patterns and properties demonstrated in a subset of the topologies are often representative for all the topologies.

This leads to a key hypothesis: we posit that \emph{if a cluster setup can violate an intent, then it can do so at relatively small scale.}  This means that it will be sufficient to verify the cluster at small scale, even if the intended cluster size is large and can vary across many dimensions.  If we cannot find a failure at small scale, we can conclude that, with high confidence, there are no violations at any scale. 

Of course, it is possible to invent cluster setups that do not satisfy the hypothesis.  The question is whether it is valid \mbox{empirically} -- and if so, what scale is typically enough to ensure finding a violation?  To answer that question, we evaluated the failure cases shown in Table~\ref{tab:failure_case_1}. We use our verification workflow to explore all the cluster topologies to find the minimum scale in which the problems occur for each case, in the ``Min Violation Scale'' column. Indeed, all cases show a violation at relatively small scale, with the maximum minimum violation scale being 3 nodes and 6 pods. (We discuss the study in detail in \S\ref{sec:empircial_study}.)

The above intuition leads to our scaling algorithm: start at small scale, and test out to a size that is empirically sufficient.  In addition to performance, there is a usability reason to do this: smaller counterexamples will generally be easier for users to understand.  We discuss how we use our intuition in \sysname{} in the following section.
 
\vspace{-1em}
\subsubsection{Incremental Scaling Algorithm} \label{sec:alg}
\vspace{-0.5em}

The scaling algorithm starts the verification on a cluster topology of small size and gradually increases the size until we find violations or reach a \emph{confident size}, i.e., a scale at which we have high confidence that the violation should have been found if there is one.
There are a couple of questions to answer for the scaling algorithm:  (1) How to determine the \emph{confident size}? (2) How should we scale up the cluster topology? (3) What if we have multiple node or pod types? 

To determine the \emph{confident size}, \sysname{} leverages a library of known failure cases (in our experiments, those in Table~\ref{tab:failure_case_1}) to empirically find at which minimum scale the failures have been found for all the cases. In particular, we first find a set of \emph{minimum violation sizes}, where each \emph{minimum violation size} for a particular failure case denotes the minimum size among all the topologies that the violations occur for that case. We then find the maximum size of nodes among all these minimum violation sizes and define it as the confident size for nodes (i.e., $n_{conf}$). We define the confident size for the pods 
as the ratio of the number of pods to the total number of nodes (i.e., $\theta_{conf}$), as the number of pods that a cluster can host will generally scale with the number of nodes (due to resource limits like CPU and memory). We calculate such a ratio according to the maximum value of the ratios among all the minimum violation sizes mentioned above. More formally, we define the \emph{confident size} as $Conf$: 
\vspace{-0.5em}
\[
\begin{array}{ll}
Conf = \langle n_{\text{conf}}, \theta_\text{conf}\rangle &
n_{\text{conf}} = \mathcal{N}\ (\max\limits_{\forall C} \ (\min\limits_{t \in T_V^C}(\ s_t\ )\ )\ ) \cdot 2
\\
\mathcal{N} (s_t) = \sum_{\forall |N_i| \in s_t} |N_i| &
\theta_{\text{conf}} = \max\limits_{\forall C} \ ( \Theta(\min\limits_{t \in T_V^C}(\ s_t\ )\ ) \ ) \cdot 2
\\
\mathcal{P} (s_t) = \sum_{\forall |P_i| \in s_t} |P_i| &
%\Theta (|N|, |P|) = \left\lceil \frac{|P|}{|N|} \right\rceil 
\Theta (s_t) = \left\lceil \frac{\mathcal{P} (s_t)}{\mathcal{N} (s_t)} \right\rceil
\\
\end{array}
\]
%\begin{align*}
%Conf &= \langle n_{\text{conf}}, \theta_\text{conf}\rangle  \\
%n_{\text{conf}} &= \mathcal{N}\ (\max\limits_{\forall C} \ (\min\limits_{t \in T_V^C}(\ s_t\ )\ )\ ) \cdot 2 \\
%\theta_{\text{conf}} &= \max\limits_{\forall C} \ ( \Theta(\min\limits_{t \in T_V^C}(\ s_t\ )\ ) \ ) \cdot 2 \\
%\mathcal{N} (s_t) &= \sum_{\forall |N_i| \in s_t} |N_i| \\
%\mathcal{P} (s_t) &= \sum_{\forall |P_i| \in s_t} |P_i| \\
%\Theta (|N|, |P|) &= \left\lceil \frac{|P|}{|N|} \right\rceil 
%\end{align*}
$T_V^C$ is the set of topologies that have violations for setup $C$. When comparing the two scale $s_{r}$ and $s_{q}$ where $r, q \in T^C$, $s_{r} \geq s_{q}$ if $\mathcal{N} (s_r) > \mathcal{N} (s_q)$ or $\mathcal{N} (s_r) = \mathcal{N} (s_q) \wedge \mathcal{P} (s_r) \geq \mathcal{P} (s_q)$. The $\min$ and $\max$ functions are defined on this comparison. We double both $n_{conf}$ and $\theta_{conf}$ to provide more confidence.  

Taking the failure cases in Table~\ref{tab:failure_case_1} as an example, using this methodology, we can find $n_{\text{conf}} = 6$ and $\theta_{conf} = 6$, where for $n_{\text{conf}}$ is from Case \snine{} and \sthree, and for $\theta_{conf}$ is from \hone{} and \sfour. Users can also leverage their cluster failure history and execute \sysname{}'s verification workflow without setting the bounds of the scaling algorithm (Algorithm~\ref{alg:scaling}) to find the minimum violation sizes for their failure cases and empirically customize the confident size for their own clusters.

With the determined confident size, we design our scaling algorithm as shown in Algorithm~\ref{alg:scaling}. It starts from the smallest scale, and gradually increases the scale. In particular, for each node type $i$, it explores its scale from $n_i^{min}$ to $n_i^{max}$ where
\vspace{-0.5em}
\[
\begin{array}{ll}
n_i^{min} = \max(0, n_i^l) & 
n_i^{max} = \min(n_i^u, n_{\texttt{conf}}) \\
\end{array}
\]

$n_i^l$ and $n_i^u$ are the lower and upper bound of the number of node type $i$ defined by the users. After determining the number of nodes to explore for each scale, we determine the number of pods according to the total number of nodes $|N|$. In particular, with the total number of nodes $|N|$, we explore the pod size from $p_i^{min}$ to $p_i^{max}$ where
\vspace{-0.5em}
\[
\begin{array}{ll}
p_i^{min} = \max(0, p_i^l) &
p_i^{max} = \min(p_i^u, \theta_{conf}\cdot |N|)
\end{array}
\]
where $p_i^l$ and $p_i^u$ are the lower and upper bound of the number of pod type $i$ defined by the users
%, and $\Upsilon(|N|)$ denotes the number of pods that can result in trivial failure cases for $|N|$
. If there are multiple types of nodes and pods, it explores all combinations of sizes $\Pi_{i=1}^n \{n_i^{min}, \cdots, n_i^{max} \} \times \Pi_{i=1}^n \{p_i^{min}, \cdots, p_i^{max}\}$. We skip any scale that result in trivial failure cases, meaning if that scale is not meaningful (i.e., $|N|=0$) or generates non-interesting failures (i.e., when $|P| \gg |N|$, the excessive number\footnote{User can provide us an upper bound of the ratio of pods to nodes numbers in their cluster, or we can approximate it from the resource requests of pods and available resources of nodes.} of pods cannot be scheduled onto the nodes). 
To illustrate the algorithm with a tiny example, we assume there are two types of nodes both with $n^{min} = 0$ and $n^{max}=1$ and one type of pod with $\theta_{conf} = 2$. The \algoname{} then explores the following topology scale $\langle |N_1|, |N_2|, |P_1|\rangle$ in order: $\langle 1, 0, 1 \rangle$, $\langle 0, 1, 1 \rangle$, $\langle 1, 0, 2\rangle$, $\langle 0, 1, 2 \rangle$, $\langle 1, 1, 1 \rangle$, $\langle 1, 1, 2 \rangle$, $\langle 1, 1, 3 \rangle$, $\langle 1, 1, 4 \rangle$. Each of these scale will be verified in one cycle by the verifier operator (discuss in \S\ref{sec:verifier_operator}).

%\vspace{-0.5em}
\begin{algorithm}
\footnotesize 
\caption{\algoname{}}\label{alg:scaling}
%\hspace*{\algorithmicindent} \textbf{Input:} \emph{cluster\_setup} from \ref{list1} \\
%\hspace*{\algorithmicindent} \textbf{Output:} a list of \emph{scaled\_setup}.  
\begin{algorithmic}[1]
\Procedure{Scaling}{}
\For{$s_t$ in \textbf{sort}($\Pi_{i=1}^n \{n_i^{min}, \cdots, n_i^{max} \} \times \Pi_{i=1}^n \{p_i^{min}, \cdots, p_i^{max}\}$)}
    \If {not \textbf{trivalCase}($(s_t, C)$)}
        \State \textbf{verifer}($(s_t, C)$)  \Comment{verify for the scaled setup $\langle t, C \rangle$}
    \EndIf
\EndFor
\EndProcedure
\end{algorithmic}
\end{algorithm}
%\vspace{-0.5em}

Note that the reason for exploring all combinations of the scale dimensions (up to the confident size) is that we observe in practice that violations can appear or disappear depending on the exact scale values in each dimension (\S\ref{sec:empircial_study}).  We therefore check all combinations to avoid compromising confidence, and as we will see \sysname{} is still sufficiently fast.

While our intuition works for our collected failure cases, it may not be applicable to some other properties, like the failures related to proportions or absolute values\footnote{For example, a slowly progressed problem where if one node goes down in a 3-node cluster, it is $\frac{1}{3}$ of the nodes and is noticed immediately, while for 50 nodes, it becomes $\frac{1}{50}$; or absolute value problems like an application can only go down if there are too many requests at the same time.}.  
If desired, users can use \sysname{} to check on a specific scale in a live cluster (by providing the cluster logs), a range of scale of their interests, or even the entire scale-spaces. But we don't recommend checking on the entire scale-spaces in terms of performance.

\vspace{-1.2em}
\subsection{Workflow of Verifier Operator} \label{sec:verifier_operator}
\vspace{-0.5em}

Consider again Figure~\ref{fig:veri_sys_arch}. The \algoname{} (\circled{2}) takes in the cluster setup (\circled{1}) and generates a sorted array of scaled setups. The verifier (\circled{3}) takes each scaled setup in order and generates a profile with the setup and other verification parameters. The parameters include options set by the users, e.g., stop at the first violation or find all, whether to enable randomness in search (see optimization options in \S\ref{sec:optimization}), or other internal parameters like the queue size for the controllers (see \S\ref{sec:model_overview}) or verifying for a subset of intents at a time if applicable. The profile is sent to the model generator (\circled{4}) to generate the SPIN model (\circled{5}, see \S\ref{sec:model_overview}). Then the model checker is triggered to compile and verify the model (\circled{6}) and sends the result back to the operator (\circled{7}). 

If a violation is found, the verifier can generate the counterexample by analyzing the error trace produced by SPIN, and present the result to the user (\circled{8}). Or it can continue to find more violations if the user has chosen the ``find all'' option. If no violation is found, the operator picks the next profile (e.g., larger queue size or another subset of intents if applicable, or the next scaled setup) and repeats \circled{3} - \circled{7} again.

%% file: modeling.tex
\vspace{-1.2em}
\section{Model} \label{sec:model}
\vspace{-0.5em}
In this section, we first discuss the overview of our model in \S\ref{sec:model_overview}. We then discuss a few optimization mechanisms to help improve run time performance in \S\ref{sec:optimization}. We summarize the implementation details in \S\ref{sec:impl}.

\vspace{-1em}
\subsection{Modeling Overview} \label{sec:model_overview}
\vspace{-0.2em}
It is challenging to effectively verify Kubernetes, as it has a complex and large implementation. In \sysname, our main goal is to verify the \emph{interactions} between control components as well as with events, rather than implementation details (i.e., error handling, data structures, APIs). Thus, instead of verifying source code, our model focuses on the high-level logic of the control components and captures the essential logic that affects the shared global states (i.e., the status of pods, nodes).

\sysname{} leverages the explicit state model checker SPIN~\cite{spin}. We choose SPIN because it can help to effectively verify the interactions mentioned above, as SPIN targets the efficient verification of concurrent and asynchronous distributed software. SPIN provides a high-level language PROMELA to specify system behavior as non-deterministic automata. Each asynchronous process is modeled as a \texttt{proctype} process in PROMELA, and SPIN will exhaustively explore all possible interleavings between these processes through a depth-first search (DFS) of a state graph that is constructed from the supplied PROMELA model. 

There are three main parts to model for Kubernetes: (1) the \emph{objects}, including nodes and workload resources like pods and deployments, (2) the \emph{controllers}, e.g., the scheduler and HPA, and (3) the \emph{events}, including environmental events like CPU change and operational events like maintenance.

\begin{figure}[t]
\centering
\includegraphics[width=3.25in]{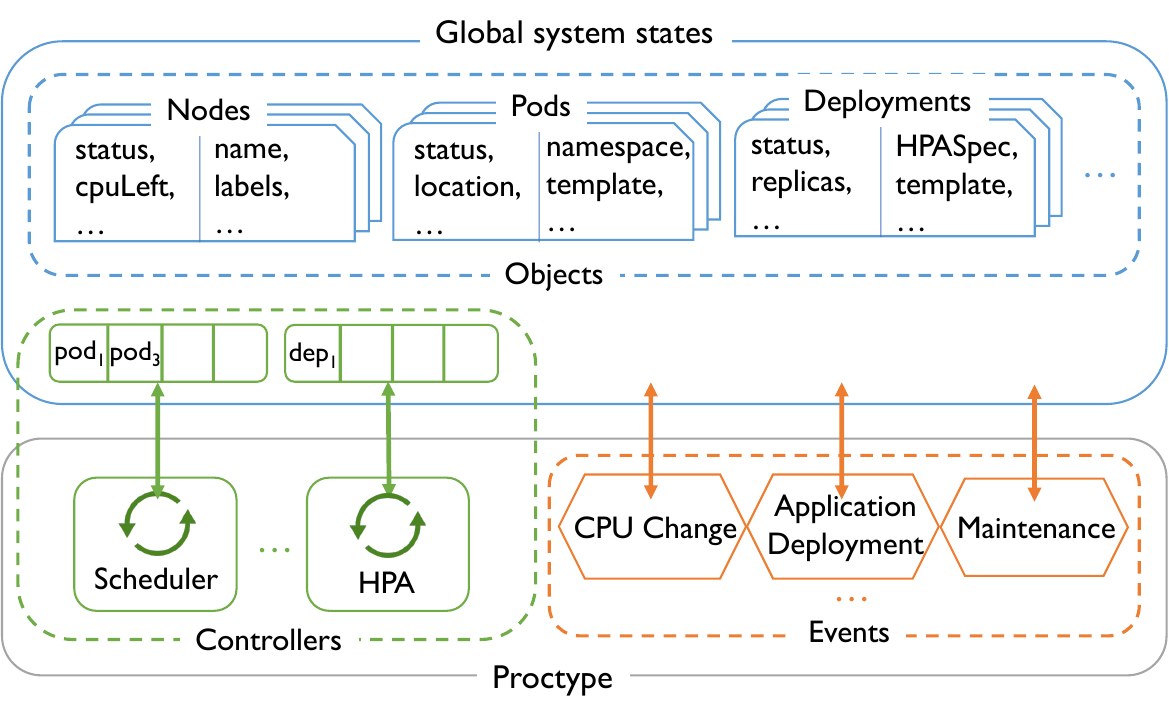}
\vspace{-1.2em}
\caption{\small \sysname{} model structure.}
\label{fig:model}
\vspace{-1.8em}
\end{figure}

As introduced in \S\ref{sec:background}, the API server provides a central place for controllers to query and manipulate the state of the objects. Our model is structured in a similar way as shown in Figure~\ref{fig:model}: we model \emph{controllers} and \emph{events} using the \texttt{proctype} in PROMELA, and model \emph{objects} using global system state that is shared across all the \texttt{proctypes}. The design differs from Kubernetes's API-style interaction slightly -- instead of listening to the updates from the API server, the controllers trigger each other. This avoids having a centralized procedure that needs to update its state after each control loop, which can unnecessarily add search depth and affect verification optimization like partial order reduction.

\myparab{Modeling the objects.} We model each kind of object using an array of customized \texttt{typedef} in the global states. Each object contains a set of attributes that are related to the system behaviors of interest. We model the intermediate states of the objects, for example, the pending state for pods. Listing~\ref{listing:1} shows a code snippet of a node definition. 

\noindent\begin{minipage}{.18\textwidth} 
\begin{lstlisting}[caption=Defining nodes.,frame=tlrb, label={listing:1}, language=c]
typedef nodeType{
    short status;
    short cpuLeft;
    short numPod;
    ...
}
nodeType nodes[SIZE];
...
\end{lstlisting}
\end{minipage} 
\hspace{.025\textwidth}
\begin{minipage}{.26\textwidth}
\begin{lstlisting}[caption=Defining scheduler.,frame=tlrb, label={listing:2}, language=c]{Name}
byte sQueue[MAX_SCHED_QUEUE];
short sTail, sIndex;
proctype scheduler() {
  atomic{
    do
        :: (sTail != sIndex) ->
        // control loop logic
    od;
  }
}
\end{lstlisting}
\end{minipage}
\vspace{-1em}

\myparab{Modeling the controllers.}
We model each controller in an event-driven loop. The queue is in the global system states that can be enqueued by other controllers and events when a control loop should be triggered. For example, when the deployment controller adds a new pod, it enqueues the pod into the scheduler's event queue, and a scheduling loop will be triggered. Listing~\ref{listing:2} shows a code snippet for the scheduler.

\myparab{Modeling events.} 
We model events in \texttt{proctype}, the same as controllers, so that SPIN can exhaustively explore the interleavings of the events and controllers in non-deterministic ways. 

\myparab{Modeling the properties.} We implement the properties introduced in \S\ref{sec:properties}. Each property is either implemented as an \texttt{assert} or as a \texttt{proctype} if it involves checking a sequence of steps. Each \texttt{proctype} is implemented as a ``monitor'' process that runs independently to check for property violations. They keep their own states that track changes in the global states and use assertions to catch violations. These \texttt{proctype}s can be explored by SPIN before and after each step.

\myparab{Modeling time.} Time is an important variable in Kubernetes. For example, HPA runs periodically (every 15 seconds by default), and CPU usage of a pod changes after the initialization phase of a certain time. Modeling time can help avoid generating non-interesting counterexamples (i.e., failures caused by HPA reacting too fast), and further help to avoid unnecessary interleavings between \texttt{proctype} to improve runtime performance. SPIN does not provide a built-in notion of time. We model time as a variable $\mathcal{T}$ in the global system state. Each \texttt{proctype} $i$ has its own local time variable $\tau_i$ representing the last time of execution, and it can only execute its next control loop or event logic when $\tau_i + \delta_i \leq \mathcal{T}$, where $\delta_i$ denotes the time interval between control loops or events. $\mathcal{T}$ is updated accordingly after each execution of an event or a control loop. 

\begin{table*} \label{tab:impl}
\renewcommand*{\arraystretch}{1}
\begin{center}
\scalebox{0.8}{
\begin{tabular}{ llcc }
\toprule
\textbf{Controllers} & \textbf{Features/Plugins Beyond Basic Framework} & \textbf{Source code?} & \textbf{LoC} \\
\midrule 
Deployment/Replicaset Controller & ReCreate, RollingUpdates & \ding{108} & 199\\
\addlinespace[.3ex]
\hdashline[1pt/1pt]
\addlinespace[.3ex]
Scheduler & NodeName, NodeAffinity, TaintToleration, NodeResourcesFit, PodTopologySpread & \ding{108} & 783 \\
\addlinespace[.3ex]
\hdashline[1pt/1pt]
\addlinespace[.3ex]
Descheduler & RemovePodsViolatingTopologySpread, RemoveDuplicates & \ding{108} & 471 \\
\addlinespace[.3ex]
\hdashline[1pt/1pt]
\addlinespace[.3ex]
HPA & Metric type Utilization and Values & \ding{108} & 222 \\
\addlinespace[.3ex]
\hdashline[1pt/1pt]
\addlinespace[.3ex]
Kubelet & N/A & \emptycirc[0.8ex] & 90\\
\addlinespace[.3ex]
\hdashline[1pt/1pt]
\addlinespace[.3ex]
Node Controller & Taint Manager & \halfcirc[0.8ex] & 86 \\
%\bottomrule
\addlinespace[.3ex]
\toprule
\textbf{Events} & \textbf{Description} & & \textbf{LoC} \\
\midrule 
CPU Change/CPU Pattern Change & The CPU usage of the pods can change randomly or in a pre-defined pattern. & & 86 \\
\addlinespace[.3ex]
\hdashline[1pt/1pt]
\addlinespace[.3ex]
Kernel panic & High resource usage can cause kernel panic and node become unhealthy. & & 25 \\
\addlinespace[.3ex]
\hdashline[1pt/1pt]
\addlinespace[.3ex]
Node Failure & Node can fail non-deterministically. & & 6 \\
\addlinespace[.3ex]
\hdashline[1pt/1pt]
\addlinespace[.3ex]
Apply/Create Deployment & Users deploy their deployment configured in YAML files. & & 126 \\
\addlinespace[.3ex]
\hdashline[1pt/1pt]
\addlinespace[.3ex]
Scale Deployment & Users scale up their deployment on the fly.  & & 11 \\
\addlinespace[.3ex]
\hdashline[1pt/1pt]
\addlinespace[.3ex]
Maintenance & Users put down the nodes for maintenance and put them back when updates are done. & & 37 \\\bottomrule

\end{tabular}
}
\end{center}
\vspace{-1.7em}
\caption{\label{tab:impl} \small Kivi Implementation of controllers and events. We have implemented a subset of features or plugins beyond the basic framework for each controller. We label whether we derived the model from the source code (or from documentation otherwise). We label the line of code (LoC) which includes log printing and excludes blank lines and comments. They all share a utility library of $311$ LoC.  }
\vspace{-1.5em}
\end{table*}

\myparab{Modeling quantitative values.} Many controllers use integers or real numbers, e.g., when calculating the average CPU usage in HPA and when calculating scores in the scheduler. There are also strings like labels on nodes. SPIN can model integers yet not real numbers or strings. We pre-process and discretize any string into integers. For real numbers, we observe that they mostly appear either in configurations with two decimals (e.g., average CPU utilization threshold in HPA) or in division. We hence convert each real number $r$ into $\lfloor r \cdot 100 \rfloor$. For division, luckily, the results of most calculations are integers (e.g., replica proposals in HPA).

\vspace{-1em}
\subsection{Optimization} \label{sec:optimization}
\vspace{-0.2em}

SPIN implements DFS to search for violations. It stores the visited state spaces to implement strategies (e.g., partial order reduction) to reduce its search space. We summarize three major aspects that can affect the run time: (1) \emph{the size of the global states}, where large global states can cause huge memory usage during search and lead to potential out-of-memory and memory swapping;
%\footnote{Processes can swapped out inactive pages from memory into disk when the memory is full.}; 
(2) \emph{concurrency}, where if there are too many processes that can be chosen from for each search step, the search spaces can be huge; (3) \emph{search depth}, where large search depth can lead to a lot more search spaces. 

We discuss a few heuristic mechanisms in \sysname{} design to improve run time performance according to the three aspects.

\myparab{Clearly defining small sets of mutating global variables.} 
To reduce the state size, we carefully pick the attributes that are related to the properties of interest. We divide these global states into two sections for performance\footnote{SPIN stores tracked global states in its search stack. Unnecessarily storing and mutating these states can result in increased memory and time.}: \emph{a stable section} that will not be changed at runtime and is not tracked by SPIN, like pod templates, node labels and names, HPA specification; and \emph{a mutating section} that keeps changes and is tracked by SPIN, including status, resource usage, pod locations, etc.

%\vspace{-.2em}
\myparab{Reducing concurrency.}
Some controllers are deployed as one instance per object, e.g., each node has a Kubelet controller, and each deployment has its own HPA instance if enabled. The size of the concurrent controller instances can increase relative to the number of objects and the search space can explode exponentially. We instead model these multiple instances of a controller as a single process. In most cases, it is safe to model in this way, as each instance operates on its own object. In addition, we model each control loop into an atomic block to avoid interleaving between other controllers in the middle of its execution with the observation that the execution time of one control loop is negligible. 

\myparab{Reducing search steps.} While in Kubernetes some controllers (e.g., scheduler) are implemented with queuing designs and some are not (e.g., HPA runs periodically), we modeled all of them using an event-driven schema to avoid unnecessary search when controllers are not involved, e.g., HPA would not be triggered until there are resource changes. We also reduce the execution of back-to-back control loops or events into a single event, e.g., if multiple pods change their CPU back-to-back, we change all of them at the same time. 

Additionally, the default way in SPIN to check on the oscillation properties is often slow, because it only considers a loop as when all the global variables appear the same repeatedly, which is unnecessarily strict. We instead only check if the loop has appeared in key relevant variables. However, as we are not checking full system state, it is possible that this subset of state could recur without causing permanent oscillation since other parts of the system state may evolve.  Thus, to weed out most such false positives, we look for multiple occurrences. For example, when checking on the eviction and scheduling cycle, we check if an eviction flag and a deploy flag for a deployment appear in turns 3 times\footnote{One can increase this number to gain confidence. Here we believe it is worth alerting users if events repeat 3 times, even if the loop is not infinite.} in a row. 

\myparab{More options.} SPIN provides additional options to improve its runtime, e.g., partial order reduction that is enabled by default, searching up to a bounded depth and time, state compression options (e.g. leveraging bloom filter),
%to trade off coverage for less memory and potentially faster run time, 
and searching with different random seeds. Users can turn on these options through \sysname{}.

\vspace{-1em}
\subsection{Implementation} \label{sec:impl}
\vspace{-0.2em}

We carefully pick the most commonly used controllers and their features of Kubernetes. Five of them are built-in  controllers and one is an add-on controller (the descheduler)\footnote{We focus on Kubernetes v1.26.0 and Descheduler v0.27.1. Our code is publicly available on Github~\cite{kivi}.}. 
For each controller, we understand all the details of the source code and try to only model the most essential details needed to accurately capture the high-level behavior of the controllers  (e.g., the interaction between controllers). For example, we omitted error handling, retries, and handler registration. 
%To achieve high accuracy, we model the frameworks and selected features of most controllers according to their source code. 
Some controllers (e.g., Kubelet) have too many low-level details (e.g., manage the pod image) that are unrelated to the properties of interest while their higher-level functionalities are simple, so we model them according to their documentations~\cite{kubelet, node_controller}. We also implement a few common events that may result in interesting failure cases. Table~\ref{tab:impl} shows the controllers and events that we have modeled.

Our current implementation of properties includes 7 checks spanning all the four categories of \S\ref{sec:properties} (oscillation, unexpected object topology, numbers, and lifecycle). In practice, these categories may involve an expanding set of checks.
%for various situations of interest. 
For now, we have implemented a selection of checks of properties related to pods, such as \texttt{checkBalanceNode($k$)} which ensures the pods in a deployment are balanced across nodes with skew of no more than $k$, \texttt{checkExpReplicas($k$)} which checks if the number of pods is $\geq k$.
%, and others of a similar flavor. 
Though modeling other objects (i.e., nodes and traffic) can leverage the same modeling framework as pods, we leave it as future work.

%% file: eval.tex
\vspace{-1.2em}
\section{Evaluation}
\vspace{-0.5em}

We seek to answer four main questions in our evaluation: (1) Can we use realistic failures to validate our intuition and scaling approach in \S\ref{sec:small_scale}?
(2) What is the performance and scalability of \sysname{}? (3) How accurate is \sysname? (4) Can \sysname{} find new problems in Kubernetes?

To answer these questions, we evaluate \sysname{} on a test suite with the 8 realistic failure cases as shown in Table~\ref{tab:failure_case_1}. These cases are representative in terms of the involved controllers (and their features) and events, the properties (covering all four summarized in \S\ref{sec:properties}), and the type of failure reasons (covering all three as summarized in \S\ref{sec:motivating_examples}). 

To be able to evaluate the performance of our system, we need to generate test suites of various sizes. All these 8 failure cases are either discussed using a fixed size of cluster or are vague on details of their configurations. We fill in reasonable details to make each example complete and parameterized to scale up while preserving its failure pattern. To affirm our understanding of each case, we have reproduced all these cases in a Kubernetes cluster.

To be able to evaluate \sysname{} under situations both with and without property violations, we further extend our test suite by generating non-violation configurations for each failure case. We make small changes to the configurations to keep the main skeleton the same while avoiding failures. 

We perform all experiments on a 2021 Macbook Pro with an M1 Max processor and 64GB RAM and reproduction on VM having 8 CPU cores of Intel(R) Xeon(R) CPU E5-2630L 0 @ 2.00GHz and 8GM RAM.

\myparab{Failure Case Reproduction.}
We leverage Kind~\cite{kind}, a Kubernetes emulation tool, to build a cluster with adjustable size (i.e., 1 master and 1-3 worker nodes). We have successfully reproduced 7 of the 8 failure cases, but skip \sfour{} as it involves kernel panic that is hard to emulate. We used the reproduction to further affirm our scaling observation (\S\ref{sec:empircial_study}) and evaluate \sysname{}'s \correct{} (\S\ref{sec:accuracy}).

\input{heatmap}

\vspace{-1.2em}
\subsection{Empirical Study} \label{sec:empircial_study}
\vspace{-.2em}
In \S\ref{sec:alg}, we introduced the methodology to find the confident size for the \algoname{}. In this section, we demonstrate how we use the collected failure cases in Table~\ref{tab:failure_case_1} to empirically find the confident size for our evaluation. We further discuss how our study affirms our intuition in \S\ref{sec:small_scale_insight}.

We run the verification workflow with the \algoname{} without bounding it by the confident size: we explore any combination of sizes that do not have trivial failures (i.e., $|N| = 0$ or $|P| \gg |N|$). The last two columns in Table~\ref{tab:failure_case_1} show the minimum violation size for each failure case and whether such size is smaller than the size in the original failure reports. Our results show that intent violations consistently appear at relatively small scale: the maximum (across our test cases) of the minimum size needed to demonstrate a violation is only 3 nodes and 6 pods. Indeed, in 6 of 7 test cases, \sysname{} found violations at even smaller scale than the scale in the original problem report, and we confirmed these by running the configurations in a real Kubernetes cluster. This also demonstrates \sysname's ability to minimize counterexample sizes in order to simplify debugging.  Based on this result, doubling to provide additional confidence, we set $n_{conf} = 3\cdot 2 = 6$ (extracted from \sthree{} and \snine{}) and $\theta_{conf} = 3\cdot 2 = 6$ (from \hone{} and \sfour{}) for our evaluation.

To further understand how failures appear, we sweep a range of combinations of node and pod counts for two representative cases with diverse patterns in Figure~\ref{fig:heatmap_all}. Each cell in the heatmaps is colored based on the result at that scale: \emph{violation}, \emph{non-violation}, or \emph{trivial failure}. From these graphs, we make a few observations:  (1) Violations consistently appear for sufficiently large $|N|$ and $|P|$, again affirming that we can use small scale to find violations that would appear large scale. (2) The exact combinations of the $|N_i|$ and $|P|$ dimensions matter.  There are many different patterns -- generally more violations at larger scale, some clear relationships between $|N|$ and $|P|$,
but also complex patterns can emerge.  This justifies the choice in our incremental scaling algorithm (\S\ref{sec:small_scale}) to explore all combinations up to the confident size.

\begin{figure*}
\centering
\begin{minipage}{.33\textwidth}
  \centering
    \includegraphics[width=\textwidth]{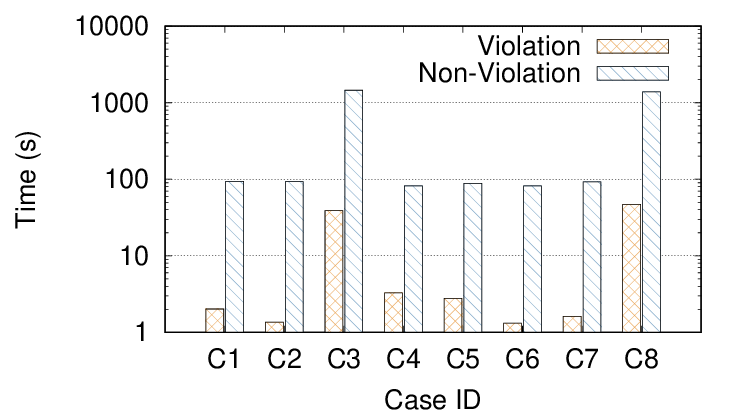}
    \vspace{-2.2em}
    \caption{\small Kivi performance. Most cases can finish within $100s$. The run time is proportional to the number of scaled setups.} \label{fig:default_mode}
\end{minipage}
\hspace{0.02in}
\begin{minipage}{.32\textwidth}
  \centering
    \includegraphics[width=\textwidth]{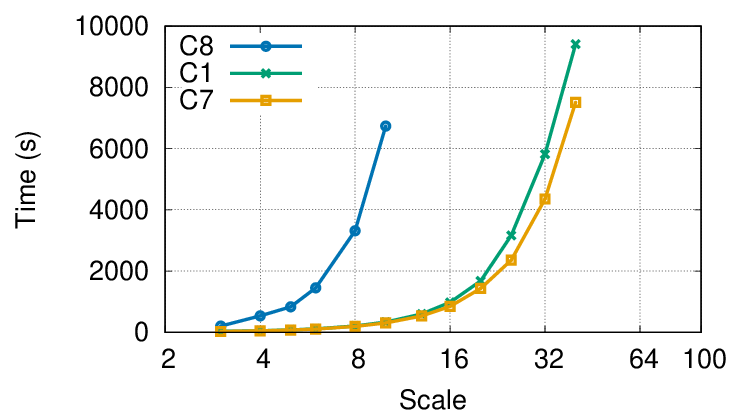}
    \vspace{-2.2em}
    \caption{\small Performance without scaling algorithm. Times out at medium sizes. (\emph{Note: this is not the actual performance for \sysname{}.})} \label{fig:wo_scaling}
\end{minipage}
\hspace{0.02in}
\begin{minipage}{.32\textwidth}
  \centering
  \includegraphics[width=\textwidth]{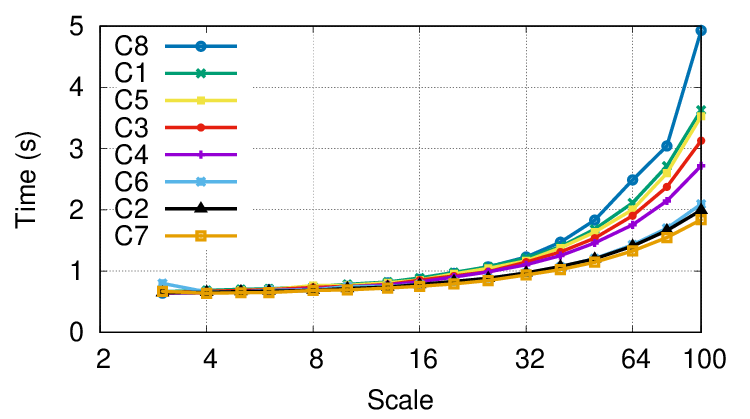}
  \vspace{-2.2em}
    \caption{\small Internal test on the model performance: non-violation cases range from $|N|=3$ to $|N| = 100$. } \label{fig:performance}
\end{minipage}%
\vspace{-0.2in}
\end{figure*}

\vspace{-1.3em}
\subsection{What is the Performance of Kivi?}
\vspace{-.2em}

Figure~\ref{fig:default_mode} shows the performance for \sysname{} under both violations and non-violations for the test suite. The result shows that \sysname verifies most cases within 100 seconds and all cases within 25 minutes. Some non-violation cases take longer, as the run time is proportional to the number of scaled setups that need to be tested: if a case contains various sizes of objects (i.e. \snine{} and \sthree{} contain $1764$ scaled setups) or is being tested for non-violation where the verification needs to explore all scale, it can take longer. Figure~\ref{fig:wo_scaling} shows the performance without the \algoname{} for three representative cases: the verification times out ($>10000$ sec) even at moderate scale ($\leq 50$ nodes).

To provide a sense of the performance of the underlying model itself, we do an internal measurement on the non-violation version of the test suite at a few specific scale ranging from $|N|=3$ to $|N|=100$, where $100$ is a large enough scale for most clusters. Figure~\ref{fig:performance} shows our model performs well even at large scale. Note, however, this shows performance for a single run of SPIN (a single topology) at each scale, rather than checking a whole range of combinations of scale parameters which we require for high coverage.

\vspace{-1.2em}
\subsection{Is \sysname accurate?} \label{sec:accuracy}
\vspace{-.2em}
We evaluate \sysname{} on both violation or non-violation versions of the test suite. \sysname{} successfully found the correct violation for all the violation cases, while reporting no failures for all the non-violation cases. 

To further evaluate the accuracy of our model in terms of controller interactions, we compare the counterexamples \sysname{} generated against the real Kubernetes event logs and see if the sequences of actions (i.e., a control action, an event) match with each other. In particular, we convert each action in the related reproduction logs\footnote{We skip the events that we don't model, like pulling an image in Kubelet, HPA failure due to unavailable metric server at bootstrapping phase, etc.} and verification counterexamples into a canonical representation.
We calculate the matching rate as the number of matched reproduction actions plus the number of matched verifier actions divided by the total number of both actions. Table~\ref{tab:accuracy} shows the results. % for our test suites. 

\vspace{-0.5em}
\begin{table}[h]
\renewcommand*{\arraystretch}{1.1}
\begin{center}
\scalebox{0.8}{
\begin{tabular}{ @{}lccc|c@{} } \toprule
  & \multicolumn{3}{P{1.4in}}{Non-deterministic cases} & \multicolumn{1}{P{1.1in}}{Deterministic cases} \\
  \cmidrule{2-4}\cmidrule{5-5}
\textbf{Case ID} &  C1 & C2 & C6 & C3, C5, C7, C8  \\\midrule
\textbf{Matching rate} & $81.6\%$  & $97.8\%$ & $100\%$ & $100\%$ \\\bottomrule
\end{tabular}
}
\end{center}
\vspace{-1.5em}
\caption{\label{tab:accuracy} \small Actions matching rates.}
\vspace{-1em}
\end{table}

Among all the cases, the cases with only deterministic events (\sthree{}, \sone{}, \done{} and \snine) match $100\%$. The other cases contain non-deterministic events: \hone{} involves CPU changes, \htwo{} and \ssix{} involve operational events. Among them, matching rates of \htwo{} and \ssix{} are near or at $100\%$\footnote{\htwo{} has one event mismatched because the verifier deleted a pod in a different node than reproduction logs, though the two nodes are symmetric.} as the operational events happen at slower speeds and don't interact much with the controllers. For \hone, $43.8\%$ of the mismatches are due to extra CPU change events in the reproduction logs. The rest are due to an inaccuracy in HPA modeling, where we do not model stabilization windows when the HPA controller is scaling down, causing the verification logs to scale down faster than the reproduction, though the final stable number is the same. Note that for non-deterministic cases in general, the accuracy as evaluated here is pessimistic because \sysname{}'s reported event sequence could be entirely valid but not the one that happened to occur in our runs of Kubernetes.

These results show good accuracy of \sysname{} in verifying realistic failure cases and modeling the interactions of controllers.

\vspace{-1.1em}
\subsection{New Issues Found}
\vspace{-.2em}
Using \sysname, we found two new issues in the implementation of the Kubernetes descheduler.

\myparab{RemoveTopologySpreadConstraint does not consider all constraints together and can mistakenly evict pods.} When there are multiple \texttt{PodTopologySpread} constraints, the \texttt{RemoveTopologySpreadConstraint} plugin decides which pods to evict per constraint instead of solving all the constraints together. This results in a sub-optimal decision to evict more pods. \snine{} is a failure caused by this issue. If implemented correctly, \texttt{RemoveTopologySpreadConstraint} should have known that the two constraints in \snine{} cannot both be satisfied and hence decide not to evict any pods. 

\myparab{RemoveDuplicates does not respect node resources and can mistakenly evict pods.} When the \texttt{RemoveDuplicates} plugin decides which pods to evict, it first collects the available nodes that can serve the duplicated pods. If there are fewer available nodes, fewer pods are evicted. However, this plugin fails to consider the available resources and mistakenly counts the occupied nodes as available, resulting in more pods being evicted than desired. \sone{} is a failure caused by this issue. If the \texttt{RemoveDuplicates} has considered node resources, no pods would be evicted.

We have submitted both issues to the Kubernetes project on Github~\cite{pull_request1, pull_request2}. The first one has been confirmed by the maintainer, where they have found a similar issue in the past. The second one is awaiting a response.

%% file: heatmap.tex
\begin{figure*}[t]
\centering
\includegraphics[width=6.4in]{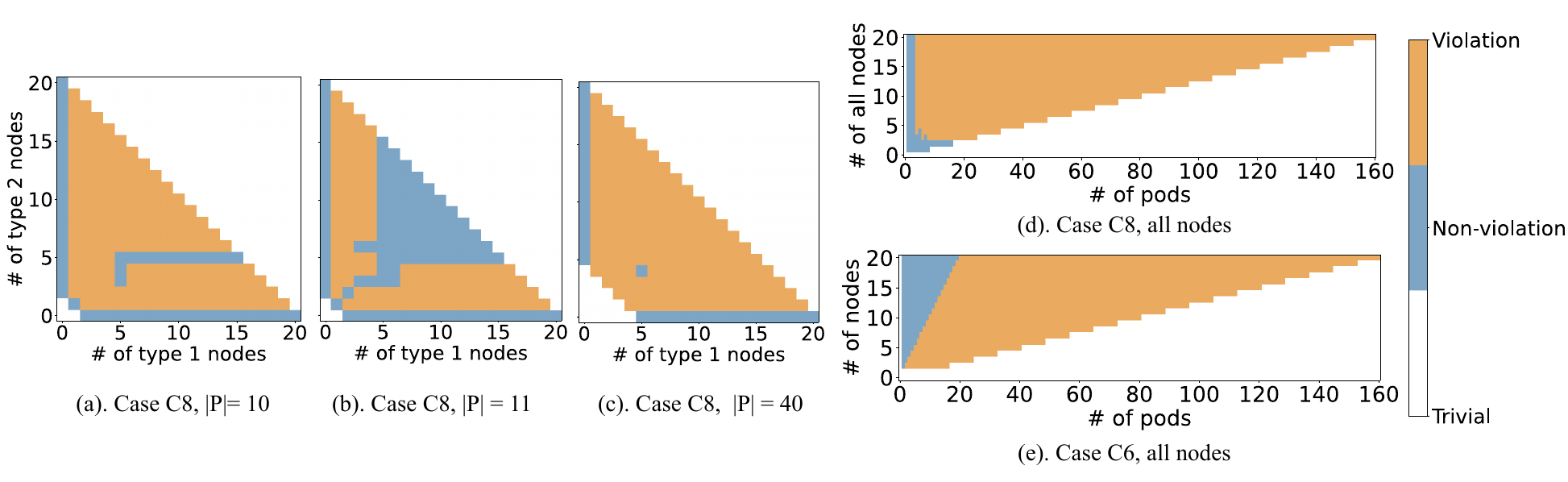}
\vspace{-1.1em}
\caption{\small Heatmap of the verification results at various scale. (e) shows the result for Case \ssix{} that scales against nodes and pods. (a)-(d) shows the result for Case \snine. \snine{} contains two types of nodes and we test on all the combinations of $|N_1|$, $|N_2|$, and $|P|$ until reaching the scale of trivial failures. (a)-(c) each shows a heatmap against $|N_1|$ and $|N_2|$ with varying $|P|$. (d) shows the result against $|N|$ and $|P|$, where if we find a violation in any ($|N_1|$, $|N_2|$), we count $|N_1|+|N_2|$ as a violation point. }
\label{fig:heatmap_all}
\vspace{-0.5em}
\end{figure*}

%% file: discussion.tex
\vspace{-1em}
\section{Discussion and Future work}
\vspace{-1em}

\myparab{Approximation.} We have made several approximations that can potentially affect the accuracy of \sysname{}.

\vspace{-0.3em}
\mypara{From Kubernetes code to model.} Our model uses fewer LoC than the original Kubernetes source. While this approximation makes verification feasible and scalable, it comes at the cost of potential approximations that may deviate from the real implementation. \S\ref{sec:optimization} and \S\ref{sec:impl} have mentioned a few such approximations. We discuss more in Appendix~\ref{sec:appro}.

With more engineering effort, our model can be brought closer to real implementations. However, there may always be some gap that is a fundamental result of verifying a model rather than verifying real code. Ultimately, Kivi is complementary to testing the real implementation; both approaches have different advantages.

\vspace{-0.3em}
\mypara{Incremental scaling algorithm.} With the scaling algorithm, \sysname may miss some failures that only manifest at large scale as discussed in \S\ref{sec:alg}. One can run \sysname{} at a specific large scale to find these issues, though there may be a performance penalty. In addition, our test suite may not be large enough to demonstrate that \sysname can empirically extract the confident size for any arbitrary Kubernetes cluster. It is in fact hard to collect real Kubernetes configurations. Hopefully, future work can build or leverage a larger dataset for empirical study.

\myparab{Limitation in Scalability and Future Work.} While \sysname{}'s \algoname{} performs well on our test setups, \sysname{} may not scale to clusters with a large degree of node and deployment type diversity, where the number of scaled setups to verify can grow exponentially with the number of types. 
However, we have found that empirically we need to verify only a small number of types. For the nodes, it is recommended that a cluster should minimize the number of node groups (i.e., set of nodes that share the same properties)
to ensure the cluster autoscaler can perform well on large clusters~\cite{node_group_1}. For the deployment, we find that most failures happened for a single deployment (i.e., most configurations are defined per deployment) and hence verifying one deployment at a time while removing the resources taken by other deployments is often enough. 

If users are still interested in operating more types of nodes or interested in the interactions between deployments, we also propose a few optimizations to explore as future work:

\vspace{-0.2em}
\mypara{(1) Divide and conquer.} For multiple deployments, we can still verify one deployment at a time, and abstract the impact of other deployments together into a small set of arbitrary ``external'' events (e.g., CPU changes on shared nodes). 

\vspace{-0.2em}
\mypara{(2) Partial order reduction on symmetric objects.} 
%We can merge any two types of objects where their differences do not affect the properties of interest. 
During model checking, we could implement an additional partial order reduction mechanism to reduce the search branches that explore the symmetric objects. 

\vspace{-0.2em}
\mypara{(3) Multi-core.} One can leverage multi-core computation to parallelize the verification for various scale.

\vspace{-0.2em}
\mypara{(4) Faster ramp up in \algoname.} Instead of increasing the size by 1 at a time to explore all combinations of sizes, we could increase it by 2 or even multiply it by 2. With the heatmap result in Figure~\ref{fig:heatmap_all}, increasing the speed can still potentially catch the failure with high confidence. We leave this exploration for future work.

%% file: related_work.tex
\vspace{-1.2em}
\section{Related Work}
\vspace{-1em}
\myparab{Verification for systems.} There are several works on applying formal methods to the domain of \domainname. Turin et al.~\cite{turin20isola} demonstrate a Kubernetes formal model yet it is based on much simplified assumptions of controllers and does not test it for verification. Liu et al.~\cite{liu2020towards} presents a proof-of-concept verification approach yet it models a couple of selected failure scenarios rather than a comprehensive implementation of controllers.
Flux~\cite{dingautomated} applies verification to serverless applications yet focuses on idempotence properties from the application perspective. Compared to these works, \sysname{} is a more comprehensive verification system based on a novel set of properties derived from real Kubernetes issues. 

Verification has been successfully applied to many distributed systems (e.g.,~\cite{zhang_et_al:LIPIcs.ITP.2021.32, DBLP:conf/osdi/HanceLHHJP20, DBLP:conf/osdi/Sigurbjarnarson18, DBLP:conf/sosp/ChajedTKZ19, ironfleet, rehearsal, DBLP:conf/sosp/ZouDDFGC19, DBLP:conf/osdi/YaseenABCL20, DBLP:conf/sosp/BornholtJACKMSS21, DBLP:conf/sosp/TaoYLLNG21, verdi}) as well as in networking (e.g., ~\cite{liveness, minesweeper, kinetic, netsmc, liveness}). While \domainname are also a distributed system, existing work is insufficient to verify them. First, many such works focus on verifying specific protocols (e.g., Paxos, BGP) instead of dynamic closed-loop controllers. Second, many of these works rely on theorem proving, which involves a great deal of human effort and that can be hard to apply to the ever-evolving ecosystem of \dname implementations. Third, the properties that \domainname{} need to verify (\S\ref{sec:properties}) are quite different from what these works focus on (e.g., crash-safety, integer overflow, network reachability). However, \sysname shares some of the underlying technologies with distributed verification work such as the use of model checking techniques~\cite{samc, modist, macemc, demeter}.
% and SPIN~\cite{spin}) in particular as a vehicle.

\myparab{Cluster management reliability.}
Several works seek to improve the reliability of \domainname~\cite{acto, sieve, aegis, lepiller2021analyzing}. Sieve~\cite{sieve} presents testing tools focusing on state reconciliation issues in customized Kubernetes controllers. H\"ayh\"a~\cite{lepiller2021analyzing} presents a tool to detect intra-update sniping vulnerabilities using dataflow graph analysis. \sysname{} targets different issues and is complementary to these works in improving various perspectives of the \domainname. 

%% file: conclusion.tex
\vspace{-.15in}
\section{Conclusion}
\vspace{-.1in}
We present \sysname{}, the first system for verifying \dname controllers and configurations. \sysname{} empirically demonstrates the insight that failures that happen at large scale can manifest at small scale and leverages it to tackle the scalability challenge. \sysname{} has shown good performance and accuracy in verifying realistic failure cases and showcases two new issues in Kubernetes controller source code.

%% file: appendix.tex
\appendix

\begin{table*}[!h]
\renewcommand*{\arraystretch}{1}
\begin{center}
\scalebox{0.8}{
\begin{tabular}{P{0.4in}P{5.2in}P{1.2in}P{1.3in} }
\toprule
\textbf{Case ID} & \textbf{Description} & \textbf{Properties} & \textbf{Reasons}  \\
\midrule 
C9~\cite{s5} & Pods are being scheduled into the same node due to the image locality plugins outweighs all other scoring plugins & \pplacement{} & \rcon{} \\
\addlinespace[.3ex]
\hdashline[1pt/1pt]
\addlinespace[.3ex]
C10~\cite{s5} & Network issues caused some pods to become unreachable (yet still healthy) and in turn their average CPU usage had dropped, causing the HPA to scale down pods and further reduce the capacity. & \pnumber{} & \revents{} \\
\addlinespace[.3ex]
\hdashline[1pt/1pt]
\addlinespace[.3ex]
C11~\cite{na1} & Nodes that were supposed to only host the App pods failed to set the right taint, causing new Daemonset pods to run on every node while App pods failed to be scheduled. & \plifecycle{} & \rcon{}  \\
\addlinespace[.3ex]
\hdashline[1pt/1pt]
\addlinespace[.3ex]
C12~\cite{outage_4} & The priorities of the pods in the production cluster were not set correctly and the deployment of a new cluster with higher priority triggered a cascading failure to preempt all the production pods. & \plifecycle{} & \revents{} \\
\addlinespace[.3ex]
\hdashline[1pt/1pt]
\addlinespace[.3ex]
C13~\cite{l1} & The ingress controller kept sending traffic to the pods that are pending deletion by the deployment controller. & \plifecycle{} & \rcontroller{}  \\
\addlinespace[.3ex]
\hdashline[1pt/1pt]
\addlinespace[.3ex]
C14~\cite{s5} & The scheduler scheduled more pods on one node over the other, while the ingress controller scheduled the traffic randomly. This caused more traffic imbalance across nodes. & \pplacement{} & \rcontroller{}  \\
\addlinespace[.3ex]
\hdashline[1pt/1pt]
\addlinespace[.3ex]
C15~\cite{na1} & One availability zone (AZ) had poor node availability, causing the node autoscaler to scale up nodes in another AZ. Later the nodes came back, causing the autoscaler to scale down the newly created nodes and pods failed to be scheduled due to bounded volumes to deleted nodes. & \plifecycle{} & \revents{}  \\
\addlinespace[.3ex]
\hdashline[1pt/1pt]
\addlinespace[.3ex]
C16~\cite{d4} & When a system component became unavailable, the logging components all woke up and cumulatively consumed high resources, causing the nodes to report unhealthy, at which time the scheduler moved the workload to the healthy node. The previous unhealthy nodes became healthy as the workload was moved while the previous healthy node became unhealthy, and the cycle perpetuated. &  Oscillation / Unexpected
Object Lifecycle (object is the controller)   & \revents{}  \\
\hline

\end{tabular}
}
\end{center}
%\vspace{-1.5em}
\caption{\label{tab:failure_case_2} \small Extended Failure Cases.}
\end{table*}
\section{More failure cases}\label{sec:more_failure}
We list more failure cases in the extended Table~\ref{tab:failure_case_2}.

\section{Approximation in \sysname{}} \label{sec:appro}
We summarize a list of major approximations when implementing the model.
\begin{itemize}[leftmargin=*]
    \itemsep0em 
    \item We do not model the workload (i.e., user requests) and we abstract the impact of these workloads into resource changes (i.e., CPU resource usage change on pods). If users are interested in the properties of the workload (e.g., if traffic distribution is unbalanced), we currently cannot support these intents. 
    \item  We only model subsets of the features in Kubernetes as listed in Table~\ref{tab:impl}. If a cluster includes unmodeled features, our model may not provide accurate results. Examples include pod graceful termination period, pod priority, pod disruption budgets, HPA stabilization windows, and StatefulSets. 
    \item We do not model anything related to the container image or smaller elements than a pod, e.g., at the container level. 
    \item We put one control loop into an atomic step and merge back-to-back control loops and events. Normally, one control loop or event can finish in a very short time. However, this can miss failures caused by such small transiting time windows.
    \item We omit error handling. If there are failures caused by the behavior of error handling, we cannot catch them. 
    \item We approximate real numbers to two decimal places, which may lead to a loss of precision, though we have not yet seen any cases that need such precision. 
    \item We implement the oscillation properties by examining if the loop has appeared on a small set of key relevant variables. As we do not check full system states, it is possible the loop we found does not cause permanent oscillation.  
\end{itemize}